\documentstyle[preprint,aps]{revtex}
\begin{document}
\draft
\title{
BCS-Universal Ratios within the Van Hove Scenario
}
\author{R. Baquero and D. Quesada$^{\rm a}$}
\address{
Departamento de F\'{\i}sica, CINVESTAV, A. Postal 14-740, 07000 M\'exico D.F.
}
\author{C. Trallero-Giner }
\address{
Departamento de F\'{\i}sica Te\'orica, Universidad de La Habana, 10400 Habana,
Cuba
}
\maketitle
Within conventional superconductivity, BCS theory represents a
well defined weak-coupling limit. The central result are the
Universal Ratios  which do not depend on physical parameters of
the particular superconductor under study. Several attempts have
been made to introduce the van Hove Scenario within BCS theory but
in none of them the Universal Ratios of conventional superconductivity
appear to be a number independent of parameters. This fact prevents the
precise definition of a deviation from the BCS value for a particular
superconductor. This concept is at the basis of several applications of
BCS theory in characterizing conventional superconductors. We define a system
that constitutes a weak coupling limit that retains the
essential features of the high-Tc oxides and which does not differ in
any essential way from other models widely used
in generalizations of BCS theory to high-Tc superconductors. The
difference is that we found a natural way of dealing with the
mathematics of the problem so as to get
Universal Ratios in the same sense as in conventional superconductivity. This
formulation could play an analogous role as the original BCS theory played in
conventional superconductivity.
\begin{abstract}
\end{abstract}
\narrowtext

Within conventional superconductivity (CS), BCS theory describes the physics
of the Cooper pairs. In its traditional formulation a Cooper pair is a bound
state formed by two electrons at the Fermi  level with opposite crystal
momentum and spin that attract themselves via a constant potential. BCS theory
neglects several important features of CS but retains the essential ones and
represents a well defined weak-coupling formulation of the problem. BCS is a
one-parameter theory. The behavior of a particular superconductor is
determined solely by the product $N(0)V$, where $V$ is the value of the
attraction constant potential between the two electrons that constitute a
Cooper pair and $N(0)$ is the electronic density of states (DOS) at the Fermi
level, $E_F$, in the normal  state. The DOS is assumed to be constant in the
energy interval around $E_F$, where  $V$, is considered to be different from
zero. According to this
model \cite{ric}, there is a  law of corresponding states in the sense that,
if $A(T)$ is a physical quantity at a temperature $T$, then $A(T)/A(0)$ or
$A(T)/A(T_c)$ (whichever exits) is a universal function of the reduced
temperature $t \equiv T/T_c$, where $T_c$ is the critical temperature.
This law is approximately followed by conventional superconductors (CSr). A
central result of this theory are the Universal Ratios (UR). These relate
thermodynamic quantities among themselves in a universal way, {\it i. e.} they
are equal to the same number for every possible superconductor.

When compared to experiment, BCS turns out not to be an exact theory of CS.
Deviations from it are frequent in all but some of the weak coupling CSr as
Al, In, and Sn. Since it is a precisely established limit, the deviations from
it do have a well defined physical meaning. This theory is useful as a point
of reference and as a tool to study different, more complicated
physical situations in a transparent and easy way from which trends can be
established. It is important to recall that, in the spirit of the theoretical
treatment of CS, the deviations from BCS theory characterize the particular
superconductor, not the theory. The validity of BCS theory itself is based on
its approximate description of the most important experimental trends and on
its characterization as a well-defined weak
coupling limit. This is a very important point for this letter. The precise
many-body theory of the electron-phonon (e-ph) CS is described through the so
called Eliashberg gap Equations \cite{eli} which give a detailed account of
the experimental results found so far for e-ph CSr.

In this letter, we want to start with a precise definition of a weak-coupling
limit for high-Tc superconductivity (HTCS). It will not be essentially
different from the models used widely in the literature. The difference is
that we found a natural way of dealing with the mathematics of the problem so
as to get UR in the same sense as in CS.
The mathematical approximations that we use are essentially of the same kind
as the ones used by other authors. Nevertheless, we keep always in mind that
the idea is to follow very closely the original BCS formulation so as to get
universal behaviors and UR.

As a final result, we will get for the UR numbers that do not
depend on any parameter characterizing the superconductor under study allowing
the precise definition of deviations from the BCS values for any given
superconductor. The new numbers do differ from the ones predicted by the
original BCS theory. This is obviously due to the 2D formulation, {\it i. e.}
to the introduction of the van Hove singularity (vHS). We speculate about the
possible meaning of the new numbers as compared to the original ones by
relating them to the trends observed in CS for $T_c$ associated to the
symmetry of the wave function
({\it i.e.} whether it is s, p or d) for the electrons occupying energies
around the Fermi level.

At the end, we will not pretend to have a precise explanation of the copper
oxides but a precisely defined weak coupling limit of high-Tc
superconductivity in the same spirit of conventional BCS theory. Such a
formulation could play a somehow similar role in HTCS as BCS played in CS and
does not exist at present in the literature although some generalization of
BCS theory do.

Several authors \cite{labbe,dza,mark,tsuei,dellano,goi,bok,chin} have
attempted to extend BCS theory to describe the HTCS. Essentially the efforts
were directed to include the vHS in the DOS into the otherwise standard BCS
theory. These formulations lead to quantitative different results from those
obtained using a constant DOS and give an approximate explanation of most of
the experimental results obtained for HTCS. In some cases the deviations found
are considerable and leave little doubt that BCS theory is insufficient to
give a detailed explanation of HTCS. Strictly speaking this is also the case
in CS \cite{carbot,baq}.

In all these attempts one of the efforts is to get an expression for the BCS
UR. The most popular of them are
\begin{equation}
R_1 \equiv {2 \Delta(0) \over K_B T_c} ,
\label{r1}
\end{equation}

\begin{equation}
R_2 \equiv {\Delta C(T_c) \over C_n(T_c)} ,
\label{r2}
\end{equation}

and

\begin{equation}
R_3 \equiv  {H_c(0) \over \sqrt{N(0)} \Delta(0)},
\label{r3}
\end{equation}

\noindent
where $\Delta(0)$ is the superconducting energy gap per electron at T=0, $K_B$
is the Boltzmann's constant , $\Delta C(T_c) \equiv C_s(T_c)-C_n(T_c)$ is the
difference between the electronic specific heat per unit volume in the
superconducting and the normal state at $T=T_c$, $H_c(0)$ is the thermodynamic
critical magnetic field at $T=0$. In the free electron approximation, not
suitable for HTCS, $C_n(T_c)= \gamma T_c$, with $\gamma= 2 \pi^2 K_B^2
N(0)/3$. It is important to remark that the magnitude $N(0)$ has no analog in
HTCS where the strong variation of the DOS near $E_F$ is the central point. In
this sense there is no expression in HTCS that would be exactly equivalent to
$R_3$ as $R_2$ and $R_1$ are. Nevertheless we will show below that an
analogous expression is possible and can be used as a UR.

For CS, BCS predicts $R_1=3.52$, $R_2=1.43$ and $R_3=\sqrt{4\pi}$. For HTCS,
different authors have obtained that these ratios  are not
universal, they are given as functions of "reasonable values" for some
parameters. Sometimes it is not clear that these have a precise meaning
consistent with the 2D formulation as, for example, the 3D Debye frequency.
For this reason, it is not possible to define in a precise way what a
deviation from the BCS value is for a particular superconductor. This is a
very useful concept within the conventional formulation of the theory.

Furthermore, usually, research is presented with the intention either to
justify or to reject BCS theory as a detailed explanation for HTCS. This is
not the spirit in the conventional theory. Nevertheless, several authors
\cite{labbe,mark,tsuei,dellano,goi,chin}, based on the way in which they look
at the approximate account of the available experimental results that they get
from their generalization, rise a conclusion for or against BCS as if one
could pretend that it is an exact theory. There is also an undeclared trend to
associate it with the electron-phonon interaction although actually no
specification about the mechanism is really necessary.

In conclusion, the attempts done so far to extend BCS theory to HTCS have
failed in getting UR, on one side, and have been usually addressed with the
idea of presenting a full explanation of the physics of HTCS which is contrary
to the spirit of BCS theory, on the other. BCS, we emphasize, neglects from
the outset several important details, as the many-body effects of the
electron-electron attraction, and cannot, as a natural result, give account of
all the experimental details of the phenomenon of superconductivity. For these
reasons, the overall conclusion one could draw from the research done
so far in this direction  is slightly undefined. Here we make an attempt to
settle this points to some extend.

Let us now define what we call the weak-coupling limit. The scenario in which
superconductivity takes place in the copper oxides lies on the copper-oxygen
planes. We start by setting a tight-binding Hamiltonian of a plane of $Cu$ and
$O$ atoms forming a square lattice. We consider first nearest neighbors
interaction with $\tau$, the hopping parameter. The dispersion relation is
given by

\begin{equation}
\epsilon_{\vec k}=-2\tau (cos(k_xa)+cos(k_ya)).
\end{equation}

Here the $k_{x,y}$ are the components of the two dimensional reciprocal
lattice vector and $a$ is the lattice constant. We take the origin of the
energy at the Fermi level. The single spin DOS can be obtained by integrating
in the $2D$ first Brillouin zone and is equal to

\begin{eqnarray}
N(\epsilon)=\frac{1}{2\pi^2\tau}{\cal K} \left[ \sqrt{1 - { \left(
\frac{\epsilon}{4 \tau}\right) }^2} \right],\nonumber\\
\nonumber
\end{eqnarray}

\noindent
where ${\cal K}(m)$ is the elliptic integral of the first type. We will not
need any value for $\tau$ for the purpose of getting the UR. Nevertheless,
$\tau$ can be taken equal to 0.25 eV in agreement with Markiewicz \cite{mark1}
who considers that this parameter can be set anywhere from 0.25 to 0.5 eV.
This gives a bandwidth $D=8 \tau=2 eV$ which is in agreement with band
structure calculations \cite{bands} and with ARPES results \cite{arpes}. While
a more realistic model can be constructed by including the second nearest
neighbors the present model has the advantage that analytic results
can be derived. Inclusion of the above term will lead to small
modifications of the UR. The asymptotic form for ${\cal K}(m)$ in the
neighborhood of the singularity can be used to get for the single spin DOS

\begin{eqnarray}
N(\epsilon) \approx \frac{1}{4 \pi^2\tau} ln \frac{16}{\left| \epsilon
\over {4 \tau} \right| }.
\label{ene}
\end{eqnarray}

This result has been obtained by several other authors
\cite{labbe,dza,mark,tsuei,dellano,goi,bok}.We will use this approximation
below.

The gap equation can be obtained in a standard way

\begin{equation}
\Delta_{\vec k}(T)= \frac{1}{2}\nonumber\\
\sum_{\vec k'} V_{\vec k,\vec k'} \frac{\Delta_{\vec k'}}{\sqrt{\epsilon_{\vec
k'}^2+
\Delta_{\vec k'}^2(T) } } tanh \left[ \frac{\sqrt{\epsilon_{\vec
k'}^2+\Delta_{\vec k'}^2(T)}}{2K_BT} \right].\nonumber\\
\nonumber
\end{equation}

With the standard approximation for the e-e attraction potential,
\[
V_{\vec k',\vec k} = \left\{
\begin{array}{ll}
-V & {\rm if\  } |\epsilon_{\vec k}|, |\epsilon_{\vec k'}| < \epsilon_0 \\
0 & {\rm otherwise},
\end{array}
\right.
\]

\noindent
where $\epsilon_0$ is a cut-off energy that we do not associate to any
particular mechanism. The 2D-BCS gap equation follows directly

\begin{equation}
\frac{2}{V}=\int_{-\epsilon_0}^{\epsilon_0} d\epsilon \frac{N(\epsilon)}
{\sqrt{\epsilon^2+\Delta^2(T)}} tanh \left[ \frac{\sqrt{\epsilon^2+
\Delta^2(T)}}{2K_BT} \right].
\label{gap}
\end{equation}

We will show now how we do get UR. We will deal first with the gap and $T_c$.
{}From our expressions for the single spin DOS (\ref{ene}) and the equation for
the gap (\ref{gap}), we can obtain an expression for $T_c$ and $\Delta(0)$. In
deriving these equations several different approximations can be taken and
different ways lead to essentially the same result but expressed
analytically
differently. One has to bear in mind that the goal is to obtain a UR and
therefore that the expression for $\Delta(0)$ that we want has to differ
from the one for $T_c$  only by a proportionality constant. Such
expressions can be obtained quite straightforwardly to give

\begin{equation}
K_B T_c = 32 \tau exp \left( 1- \sqrt{\frac{8 \pi^2
\tau}{V}+ln^2(\frac{\epsilon_0}
{64 \tau})-1} \right),
\end{equation}

\noindent
for the critical temperature and

\begin{equation}
\Delta(0) = 64 \tau exp \left( 1- \sqrt{\frac{8 \pi^2
\tau}{V}+ln^2(\frac{\epsilon_0}
{64 \tau})-1} \right)
\end{equation}

\noindent
for the gap at zero temperature. Then

\begin{equation}
\frac{2 \Delta(0)}{K_B T_c} = 4
\end{equation}

\noindent
which is slightly higher than the BCS-value of 3.52. The difference comes
obviously from the inclusion of the vHS. We will comment on it below.

We get now the ratio $R_2$  for the specific heat jump at $T_c$ defined above
in Eq (2). We start from the expression

\begin {equation}
\Delta C(T_c)=\frac{1}{2 K_B T_c}
\frac{d\Delta^2(T)}{dT}\Bigg|_{T_c}\int_{-\epsilon_0}^{\epsilon_0} d\epsilon
N(\epsilon) f(\epsilon) (1-f(\epsilon)),
\end{equation}

\noindent
where $f(\epsilon)$ is the Fermi-Dirac distribution function. From Eq (11) we
get immediately

\begin{equation}
\frac{\Delta C(T_c)}{C_n}=\frac{3}{\pi^2 K_B^2 T_c}
\frac{d \Delta^2}{dT}\Bigg|_{T_c},
\end{equation}

Since a numerical calculation of Eq (7) gives us a behavior for the gap
function $\Delta(T)$ very similar to the conventional BCS one, we have used
the BCS value for the derivative \cite{ric}

\begin{equation}
\frac{d\Delta^2(T)}{dT}\Bigg|_{T_c}= - 9.61 K_B^2 T_c
\end{equation}

\noindent from which we get directly another UR

\begin{equation}
\Bigg|\frac{\Delta C(T_c)}{C_n}\Bigg|= 2.92
\end{equation}

\noindent which is about twice the BCS value.

The thermodynamic critical magnetic field, $H_c(0)$, can be calculated from
the 2D thermodynamic relation:

\begin{equation}
\frac{H_c^2(T)}{4\pi}=F_n(T)-F_s(T)
\end{equation}

\noindent  where $F_n\ ( F_s)$ is the free energy in the normal
(superconducting) state.

After some straightforward but slightly lengthy algebra, we arrive at the
expression

\begin{equation}
\frac{H_c^2(0)}{4\pi}= N_0 \Delta^2(0) (\frac{4}{3}-ln\frac{\Delta(0)}{64
\tau})> \frac{7 N_0 \Delta^2(0)}{3}
\end{equation}

\noindent  where we have used, $ln x<x-1$ for $x<1$ and $N_0=\frac{1}{8\pi^2
\tau}$. Therefore

\begin{equation}
\frac{H_c^2(0)}{N_0 \Delta^2(0)}>\frac{28 \pi}{3}
\end{equation}

\noindent  and the lower limit gives us again a UR

\begin{equation}
\frac{H_c(0)}{\sqrt{N_0} \Delta(0)}= 5.41.
\end{equation}

Here we have to realize that $N_0$ is not the DOS value at $E_F$ as in CS.

In conclusion, in this letter we have considered the plane of $Cu$ and $O$
atoms where superconductivity takes place in the HTCS. A tight-binding
formulation of the problem has allowed us to get the DOS in agreement with
previous authors. Since BCS theory has been so useful in CS, based on its
universality and its simplicity, we found natural to attempt a similar
formulation valid for HTCS. Our result differs from previous work. Our main
conclusion are the UR given by Eqs.(10), (14) and (18). We want now to
speculate on the reason to get higher numbers than in the conventional theory.
In conventional BCS theory, the trend of $T_c$, $\Delta(0)$ and the UR with
the behavior of the electronic structure was not usually considered since the
only electronic parameter was the DOS at $E_F$ that entered always as a
constant. A more careful look at the symmetry of the wave function of the
electrons at the Fermi level (whether they are s-, s-p or d-like) shows that
to higher values of the quantities mentioned above corresponds a more
localized (in the sense stated above) wave function. In other words, a system
with electrons described by
s-waves at $E_F$ should have a lower $T_c$ as compared to another system where
the electrons at this energy are described by d-waves. As an example, $Al$
(s-wave) has a lower $T_c$ than $Nb$ (d-wave). This has nothing to do,
obviously, with the symmetry of the order parameter. We can build the
speculation that {\it \bf the more localized the wave function is, the higher
the $T_c$, $\Delta(0)$ and the UR}. In this sense one could expect the UR for
HTCS to be higher that the corresponding ones for CS, in the sense that the
phenomenon takes place in a plane which is more localized than a 3D space,
where electrons with wave functions of p and d character are present at the
Fermi level. We repeat that this relationship is somehow speculative. The real
sharp fact is that the UR are possible to formulate in a reasonable way and
turn out be higher numbers. We expect that they will play the same role in
HTCS as they did in CS. The mathematical details of this formulation will be
published elsewhere.

\end{document}